# BLASTING PARAMETERS ALTERNATE SELECTION AS A TOOL FOR ELASTIC WAVE EFFECT AMPLIFICATION AT POTENTIALLY INSTABLE LOCATIONS WITHIN MAIN ROOF STRATA

**Witold Pytel**, **Piotr Mertuszka**

KGHM CUPRUM Ltd. Research and Development Centre, **Poland**

**ABSTRACT**

*The exploitation of copper ore deposits in Polish underground mines is performed primarily by the use of blasting technology which seems to be relatively well suited to the hardness of the local rocks and to local mining and geological conditions. Blasting works are also recognized as an active method of seismic events prevention, conducted as group strain-release blasting within potentially instable main roof strata and in pillars. A notable number of recorded dynamic events can be clearly explained by the blasting works' effects. However, these operations are presently conducted intuitively based on a trial-and-errors basis, rather than upon an intentional and scientifically justified approach. As a one of seismic hazard reduction measures, a special kind of group winning blasting conducted with appropriately selected delays between the successive face-detonations is presented in the paper. This approach is able to amplify purposely the energy of rock mass vibration within the area which has been identified by the FEM modeling as a potentially instable. As a result of the entire multi-faces blasting operation one may expect the rock mass possibly distressing manifested as a seismic event. The presented analyses have been supported by a 3D FEM dynamic modelling of multi-face blasting in different configurations of space and time coordinates.*

1. ## INTRODUCTION

KGHM Polska Miedź S.A. is a Polish mining company which operates 9 opencast and underground mines located in Poland, Canada, the USA and Chile. The three underground Polish copper mines, i.e. Lubin mine, Polkowice-Sieroszowice mine and Rudna mine, are located in the South-West part of Poland, Upper Silesia region. The basic operations of the Company are primarily limited to copper ore mining, metallic copper production and production of precious metals and other non-ferrous metals. The copper deposit on the Foresudetic Monocline is of the stratified type, occurring within sedimentary rocks. Copper sulphide accumulations form the orebody occurring within the decoloured sandstones of the Rotliegendes and sandstones, copper shales and carbonaceous Permian limestone rock.

The underground mining operations are usually associated with a high seismic activity. With the growing depth of exploitation and higher variability of rock mass and roof characteristics, the difficulties in using applied mining methods caused by increasing pressure, had been observed. The largest risks for mining operations are created by high-energy tremors, the hypocenters of which are located within the main roof strata. Those seismic tremors occur equally often in the areas which are virtually free of tectonic disturbances.

Blasting works are recognized as an active method of rockburst control in KGHM's underground mines conditions. These operations are conducted as a group strain-release blasting within potentially instable roof strata. The proposed activity has focused on utilizing the effect of interference multiplication of elastic waves induced by the rationalized group blasting as a new method of rockburst control in underground mines. Interference is a phenomenon in which two or more waves superimpose to form a resultant wave of locally greater or lower amplitude than the primary waves independently. Interference usually refers to the interaction of waves that are correlated or coherent with each other, either because they come from the same source or because they have the same or nearly the same frequency. Based on preliminary conducted research on this effect using field measurements and analytical computations it was assumed that interference of elastic waves caused by group blasting may be observed in the mine scale, and amplification of elastic waves due to group blasting may be implemented as a method for stress relieve in deformed rock mass [1][2][3].

This task's goal is to develop and implement firing scenarios for multi-face production blasting such that they increase the capability of inducing stress relief in the rock, manifested as a seismic event in the rock mass being mined. This method would improve stability control in underground workings, as well as mitigate risks associated with the dynamic effects of rock mass pressure, as compared with currently used methods.

## 2. GEOMECHANICAL ANALYSIS OF ROCK-MASS BEHAVIOR

Optimization of parameters of winning blasting conducted for group of faces, aiming for amplification of elastic wave effect, requires conducting an appropriate analytical work coupled with underground validation. The proposed activity will cover therefore: (a) selection of trial panel; (b) numerical modelling of rock mass behavior in the vicinity of trial panel (stress-strain-safety fields); (c) identification of weak/prone to instability rock mass' areas (based on margin of safety 3D distribution); (d) numerical modelling of elastic waves induced by blasting and (e) parametric analyses based on numerical modelling of group winning blasting under different conditions (geometry, number of mining faces blasted in one cycle, delays etc.).

Geomechanical problem solution and results visualization were based on the NEi/NASTRAN computer program code utilizing FEM in three dimensions [4]. It was assumed that all materials reveal linear elastic characteristics, except for rocks comprised within pillars that exhibit elastic-plastic behavior with strain softening. The entire numerical models general boundary conditions were described by displacement based relationships. More information on the applied solution method may be found elsewhere. Numerical experiments permitted determining the overall stress/deformation states which were used afterwards for quantitative characterization of the actual level of safety using the indicators called safety margins related to different failure criterions. The value (positive/negative depending on the strength hypothesis involved) of any safety margin indicates the likely occurrence of instability in the rock mass.

Taking into account that some solid rocks may fail in tension at stress of about 3.0 MPa, the corresponding particle With the specific strength parameters of a rock mass and the calculation results in the form of stress and deformation fields for the mining conditions under consideration, the safety margins for the roof rocks were determined using the following criteria [5]:

- $F_{cm}$ – **Mohr-Coulomb criterion** (maximum shear stress) which is associated with the mechanism of destruction due to shear deformation of a rock mass:

$$F_{cm} = -\sigma_3 \cdot \frac{R_c}{R_r} + R_c - \sigma_1$$

where: $R_c$ – uniaxial compressive strength of a rock mass (MPa),

$R_r$ – uniaxial tensile strength of a rock mass (MPa),

$\sigma_1, \sigma_3$ – principal stress (MPa).

- $F_{vm}$ – **Huber-Mises criterion** (reduced stress) which is associated with the mechanism of destruction due to compression of a rock mass:

$$F_{vm} = R_c - \sigma_{vm} = R_c - \frac{1}{\sqrt{2}}\sqrt{(\sigma_1 - \sigma_2)^2 + (\sigma_1 - \sigma_3)^2 + (\sigma_2 - \sigma_3)^2}$$

where: $\sigma_{vm}$ – stress reduced according to Huber-Mises's equation (MPa),

$\sigma_1, \sigma_2, \sigma_3$ – principal stress (MPa).

- $F_\tau$ – **Maximum shear strain hypothesis:**

$$F_\tau = \varepsilon_{nr} - \varepsilon_{\tau,max}$$

where: $\varepsilon_{nr}$ – assumed value of the destructive shear deformation ($\varepsilon_{nr} = 0.002$)

$\varepsilon_{\tau,max}$ – deformation value at failure:

$$\varepsilon_{\tau,max} = \frac{\gamma_{max}}{2} = max\left(\left|\frac{\varepsilon_1 - \varepsilon_2}{2}\right|, \left|\frac{\varepsilon_2 - \varepsilon_3}{2}\right|, \left|\frac{\varepsilon_3 - \varepsilon_1}{2}\right|\right)$$

where: $\varepsilon_1, \varepsilon_2, \varepsilon_3$ – principal deformations.

To illustrate the effect of mining face geometry as well as high horizontal stress involvement, G-54 mining district of Polkowice-Sieroszowice mine has been analysed using the FEM modelling formulated in three dimensions (Figure 1).

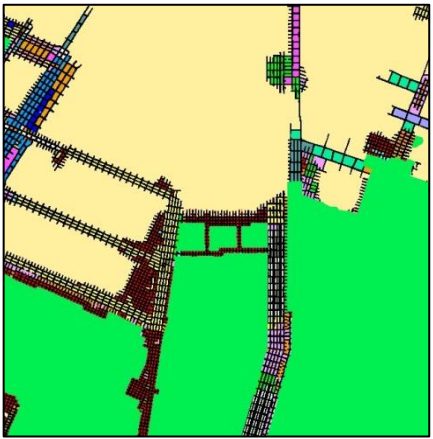 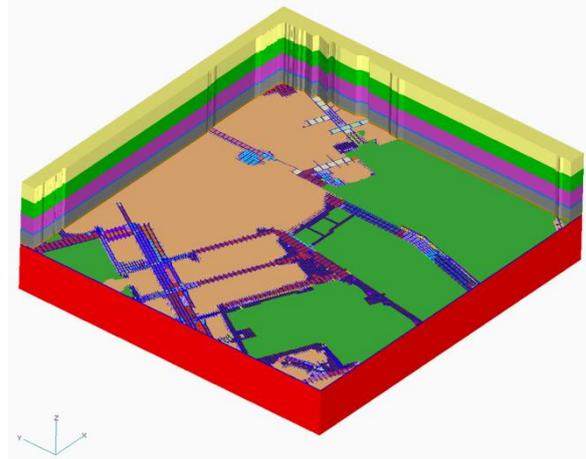

**Fig. 1. View of the FEM model of G-54 panel: ore level (left) and entire 3D model cross-section (right)**

The example of calculated safety margins' contours using the aforementioned strength hypothesis in a specific locations in the main roof strata are shown in Table 1.

**Table. 1. Calculated values of safety margin contours within roof strata based on selected strength criteria**

| | |
|---|---|
| Mohr-Coulomb criterion – 94 m above the immediate roof strata (breccia) | Mohr-Coulomb criterion – 255,5 m above the immediate roof strata (clayey shale) |
| Huber-Mises criterion – 94 m above the immediate roof strata (breccia) | Huber-Mises criterion – 255,5 m above the immediate roof strata (clayey shale) |

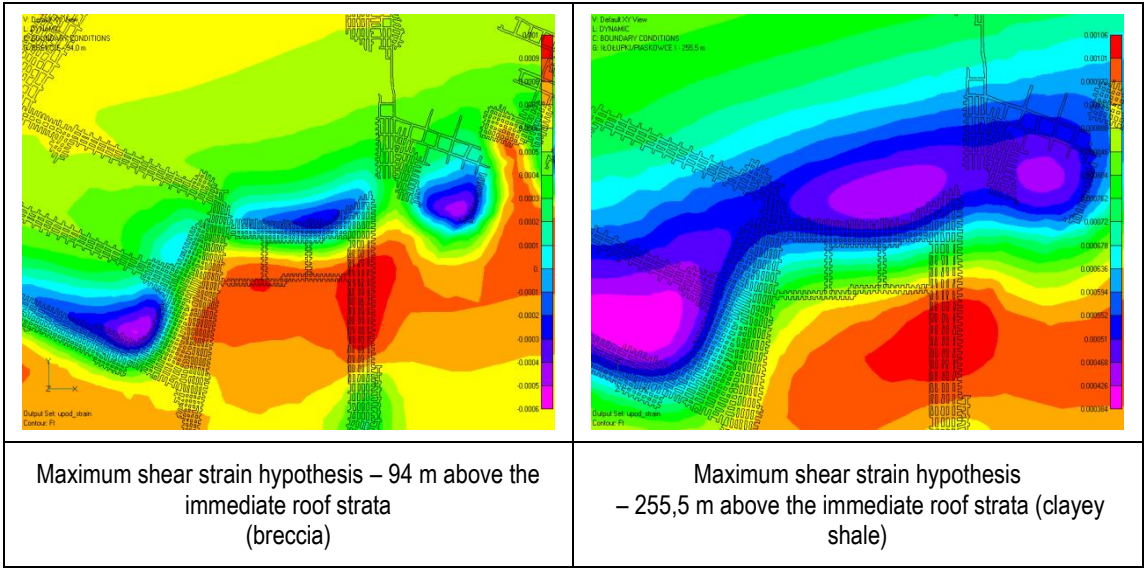

| Maximum shear strain hypothesis – 94 m above the immediate roof strata (breccia) | Maximum shear strain hypothesis – 255,5 m above the immediate roof strata (clayey shale) |

The strength calculations were performed under the assumption that the laboratory values $R_r$ and $R_c$, regarding to rock mass conditions, should be significantly scaled down according to Hoek's approach [6]. This assumption leads to a significant decrease in the safety margin actual values. The evaluation of the rockburst hazard level based on determined safety margins allows for identifying the best fitted delays' set, that is most advantageous from the safety point of view among the considered solutions. Since the best solution by assumption should induce a seismic event occurrence at the earlier identified main roof location, one would expect there the lowest as possible values of safety margins.

The analysis shows that the risk of instability events which is associated with the cracking of the roof rocks strata, e.g. due to excessive differences between the values of principal stress ($F_{vm}$), is insignificant at all roof levels close to the excavations. The high compressive strength of roof strata rocks clearly reduces this risk. The risk of instability which is associated with the possibility of roof rocks detachment (e.g. due to excessive shear stress) and represented by the $F_{cm}$ safety margin seems to be also a secondary problem at all levels above the mine workings.

Furthermore, the width of the front's opening of considered mining panel generates a hazard zone due to the roof destruction mechanism caused by high compression. This may occur due to excessive energy of elastic deformation, usually accompanied by the crushing of pillars under the vertical load transmitted from the roof strata as shown by the negative value of the $F_{vm}$ safety margin on the mining front, mainly on the right section near the goafs (mined out area) of the Rudna mine. This situation may led to rock ejections from sidewalls of the nearby pillars and/or in the crushing of pillars in this area.

However, the presented geomechanical analysis indicates that the roof instability may occur mainly in the breccia layer located at 94 m above the immediate roof strata as shown by the maximum value of the safety margin at this level, determined according to Huber-Mises criterion and the hypothesis of maximum shear strain. This kind of stress-strain conditions favour instabilities appearance within the wide areas in the breccia's plane.

The potential areas at risk should also include the shale with sandstone layer located 225.5 m above the immediate roof strata.

### 3. NUMERICAL SIMULATIONS OF GROUP BLASTING PROCEDURE

The group blasting operation performed on December 2011 within G-54 mining parcel of Polkowice-Sieroszowice mine lead to induced tremor occurrence which was registered while firing of a group of 26 mining faces. This situation was analysed in the previous chapter and indicates, that an additional load in the form of a stress change in the roof layer (due to a dynamic impulse caused by blasting operation) could lead to high-energy tremor occurrence within the identified areas of the main roof strata. Based on existing numerical model, further simulations of blasting operation in different configurations have been analysed. It has been assumed that, that the following parameters should be taken here into account: (a) the number of faces blasted per one cycle; (b) the spatial location of the mining faces and (c) delays applied between mining faces [7]. The firing of each mining face was simulated by applying hydrostatic pressure within a finite elements in which the detonation of explosives were represented by the pressure increasing from 0 to 5 GPa within 1 ms, and decreasing after detonation linearly to zero within the next 4 ms. Having in mind that the wave velocity in surrounding rock mass may reach 6,000 m/s, the applied delay time should be as short as possible. In practice, it means that seismic waves covers a distance of 6 m in 1 ms. For this purpose, electronic detonators in delay increments of 1 ms should be applied definitely. It gives a great advantage with respect to other initiation systems [8]. It was also assumed that all of the analysed mining faces were provided with an additional large-diameter hole (Figure 2) which should be filled with 30 kg of explosives and equipped with the electronic detonator

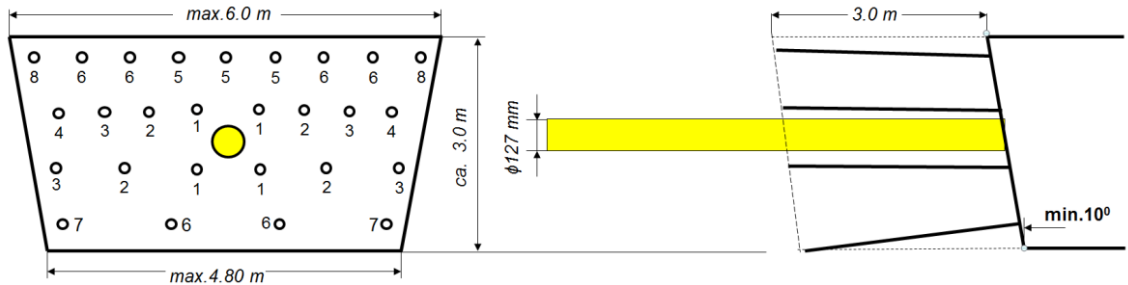

**Fig. 2. Typical drilling and firing pattern with additional hole of enlarged length and diameter**

The conducted numerical simulations of group winning blasting, involved the group of 33 mining faces. The time delays were applied based on geometrically determined distance between the analysed mining face or group of faces and the area characterized by the lowest value of safety factor calculated based on selected strength hypothesis. Changes of safety margin values in front of the faces line and 94 m above mine workings, at specific points of time within 0 to 100 ms after the firing of the faces, have been presented in the form of graphs below (Figures 3÷5).

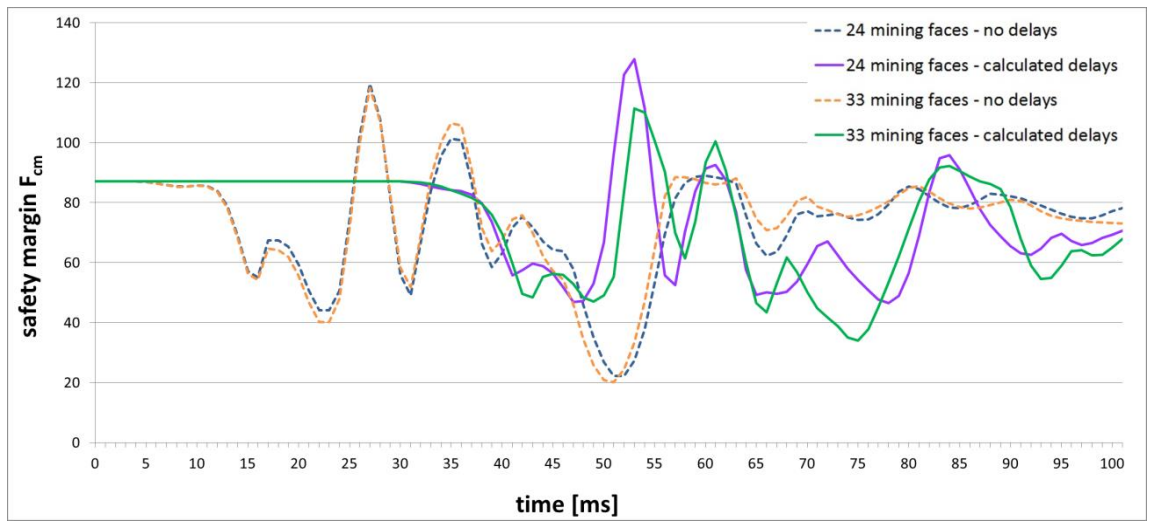

**Fig. 3. Changes of safety margin $F_{cm}$ values 94 m above the immediate roof strata**

The analysis developed within the framework of this paper have proved, that an additional load in the form of a stress change caused by blasting operations could lead to an instability within main roof strata and finally may induce the high-energy tremor. Having in mind that the negative change in the safety margin may indicate high likelihood of rock mass instability, the determined safety margins confirm an existing risk of rock mass destruction. Since both the safety margins calculated using maximum shear strain hypothesis and the maximum distortion energy hypothesis have reached negative values, the instability within the breccia stratum may be expected.

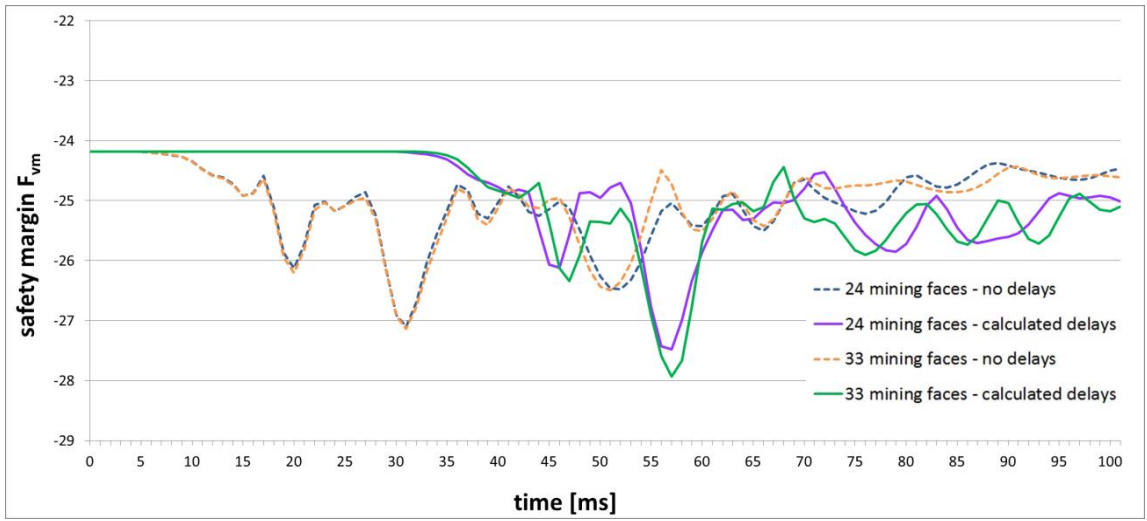

**Fig. 4.** Changes of safety margin $F_{vm}$ values 94 m above the immediate roof strata

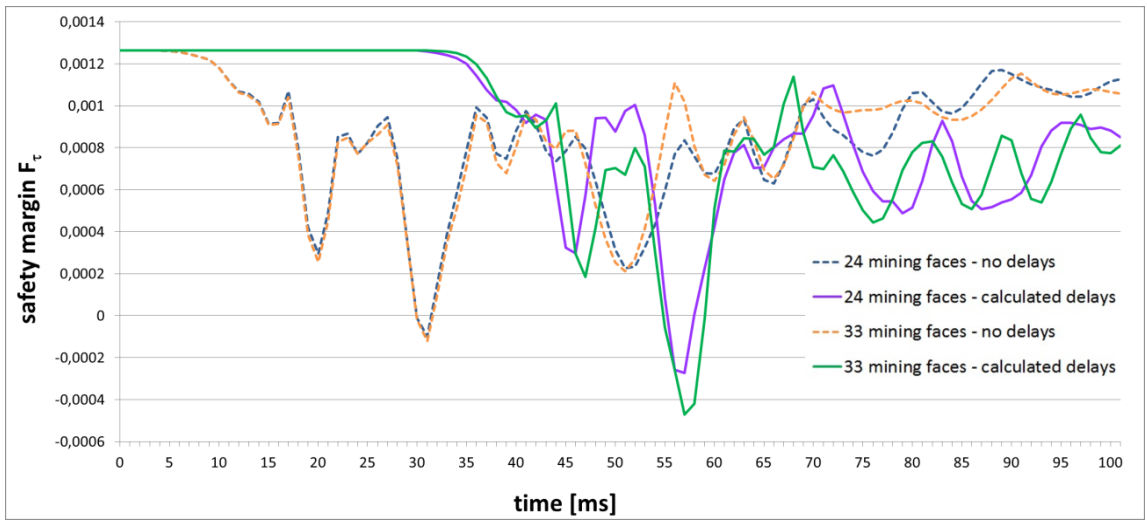

**Fig. 5.** Changes of safety margin $F_{\tau}$ values 94 m above the immediate roof strata

### 4. CONCLUSIONS

The presented results of numerical modelling have proved that there is a clear relationship between the applied delay of charges detonation and the stress-strain time-dependent changes within rock mass. The applied numerical methods have revealed themselves to be very effective in the verification of the 'primary' conditions (stress and deformations) within the analysed area of the rock mass. Changes in the number of faces, their spatial location and delays of face blasting operations per a one cycle have a considerable impact on the spatial distribution of safety margins within a rock mass. This is a very promising conclusion in respect of usefulness of the multiple faces fired with the smartly fitted mutual delays as a tool for seismic hazard reduction in mines.

The effect of group winning blasting on the effectiveness of active methods of rockbursts control still requires more research and field tests as well as more detailed numerical modelling including the effects of different model parameters. One may expect that further investigations of the detonation process, pressure wave propagation and its interaction with structures may influence the reliable results of destressing blasting simulations either. Comprehensive knowledge on the type and pressure distribution of dynamic impulse may allow for the optimal application development of presented method in blasting work designing phase, what should be confirmed by underground field tests.

## ACKNOWLEDGEMENTS

**This paper has been prepared through the Horizon 2020 EU funded project on "Sustainable Intelligent Mining Systems (SIMS)", Grant Agreement No. 730302.**